\documentclass[onecolumn,showpacs,preprintnumbers,amsmath,amssymb]{revtex4}
\usepackage{graphics,psfrag}
\usepackage{rotating}
\usepackage[usenames]{color}
\usepackage{amsmath}
\usepackage{bbm}
\usepackage{ulem} 
\newcommand{\be}{\begin{eqnarray}}
\newcommand{\ee}{\end{eqnarray}}

\newcommand{\non}{\nonumber\\}

\newcommand{\dell}{$\Delta^*(1700)$ }

\begin{document}

\title{Clues to the nature of the $\Delta^*(1700)$ resonance from pion- and photon-induced reactions}

\author{M. D\"oring}
\email{doering@ific.uv.es}
\affiliation{Departamento de F\'{\i}sica Te\'orica and IFIC,
Centro Mixto Universidad de Valencia-CSIC,\\
Institutos de
Investigaci\'on de Paterna, Aptd. 22085, 46071 Valencia, Spain}
\author{E. Oset}
\email{oset@ific.uv.es}
\affiliation{Departamento de F\'{\i}sica Te\'orica and IFIC,
Centro Mixto Universidad de Valencia-CSIC,\\
Institutos de
Investigaci\'on de Paterna, Aptd. 22085, 46071 Valencia, Spain}
\author{D. Strottman}
\email{dds@ific.uv.es}
\affiliation{Departamento de F\'{\i}sica Te\'orica and IFIC,
Centro Mixto Universidad de Valencia-CSIC,\\
Institutos de
Investigaci\'on de Paterna, Aptd. 22085, 46071 Valencia, Spain}

\begin{abstract}
We make a study of the $\pi^- p\to K^0\pi^0\Lambda$, $\pi^+ p\to K^+\pi^+\Lambda$, $K^+\bar{K}^0p$, $K^+\pi^+\Sigma^0$, $K^+\pi^0\Sigma^+$, and $\eta\pi^+ p$ reactions, in which the basic dynamics is given by the excitation of the $\Delta^*(1700)$ resonance which subsequently decays into $K\Sigma^{*}(1385)$ or $\Delta(1232)\eta$. 
In a similar way we also study the $\gamma p\to K^0\pi^+\Lambda$, $K^+\pi^-\Sigma^+$, $K^+\pi^+\Sigma^-$, $K^0\pi^0\Sigma^+$, and $\eta\pi^0 p$ related reactions.
The cross sections are proportional to the square of the coupling of $\Delta^*(1700)$ to $\Sigma^*K$ ($\Delta\eta$) for which there is no experimental information but which is provided in the context of coupled channels chiral unitary theory where the $\Delta^*(1700)$ is dynamically generated. Within present theoretical and experimental uncertainties one can claim a global qualitative agreement between theory and experiment. We provide a list of items which need to be improved in order to make further progress along these lines.
\end{abstract}
\maketitle
\section{Introduction}
The history of the dynamically generated resonances, which appear in the solution of the meson-meson or meson-baryon coupled channel Lippmann-Schwinger equation (LSE) with some interaction potential, is quite old. One of the typical examples is the $\Lambda(1405)$ resonance which appears naturally in coupled channels containing the $\pi\Sigma$ and $\overline{K}N$ channels \cite{dalitz,Jennings:1986yg}. The advent of unitary extensions of chiral perturbation theory has brought more systematics into this approach with chiral Lagrangians providing the kernel, or potential, for the LSE or its relativistic counterpart, the Bethe Salpeter equation (BSE) which is more often used. In this sense the $\Lambda(1405)$ has been revisited from this new perspective and at the same time new resonances like the $N^*(1535)$, $\Lambda(1670)$, etc. have been claimed to be also dynamically generated \cite{Kaiser:1995cy,Kaiser:1996js,angels,Nacher:1999vg,oller,Inoue:2001ip,bennhold,Garcia-Recio:2002td}. Actually, one of the surprises along these lines was the realization that the chiral theory predicted the presence of two nearby $I=0,S=-1$ poles close to the nominal $\Lambda(1405)$ mass, such that the physical resonance would be a superposition of the two states \cite{Jido:2003cb,Garcia-Recio:2003ks}. Recent work including also the effect of higher order Lagrangians in the kernel of the BSE \cite{Borasoy:2005ie,Oller:2005ig,Oller:2006jw} also find two poles in that channel, one of them with a large width \cite{Borasoy:2005ie}. Interestingly, recent measurements of the $K^-p\to\pi^0\pi^0\Sigma^0$ reaction \cite{Prakhov:2004an} show the excitation of the $\Lambda(1405)$ in the $\pi^0\Sigma^0$ invariant mass, peaking around 1420 MeV and with a smaller width than the nominal one of the PDG \cite{Eidelman:2004wy}. An analysis of this reaction and comparison with the data of \cite{Thomas:1973uh} from the $\pi^-p\to K^0\pi\Sigma$ reaction led the authors of Ref. \cite{Magas:2005vu} to conclude that the combined experimental information of these two reactions provided evidence of the existence of two $\Lambda(1405)$ states.

More recent work has extended the number of dynamically generated resonances to the low lying $3/2^-$ resonances which appear from the interaction of the octet of pseudoscalar mesons with the decuplet of baryons \cite{lutz,decu_ss}. One of the resonances that appears qualitatively as dynamically generated is the $\Lambda(1520)$, built up from $\pi\Sigma^*(1385)$ and $K\Xi^*(1530)$ channels, although the necessary coupling to the $\pi\Sigma$ and $\overline{K}N$ channels makes the picture more complicated \cite{Sarkar:2005ap,Roca:2006sz}. A more detailed discussion on the meaning of the dynamically generated resonances, its physical interpretation and the technical details on how to produce them can be seen in Sec. II of the paper \cite{Doring:2006ub}.

\begin{figure}
\includegraphics[width=6cm]{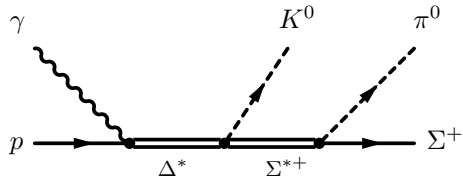}
\caption{Tree level contribution for the $\gamma p\to K^0\pi^0\Sigma^+$ reaction.}
\label{fig:tree_level_ks}
\end{figure}

Another $3/2^-$ resonance which appears in the same scheme is the $\Delta^*(1700)$, and in \cite{decu_ss} the couplings of the resonance to the coupled channels $\Delta\pi,\Sigma^*K$, and $\Delta \eta$ were calculated. The couplings $g_i$ are 1.0, 3.4, and 2.2, respectively, for these channels. It is interesting to note the large strength of the coupling to the $\Sigma^* K$ channel. Due to this, it was found in Ref. \cite{Doring:2005bx} that the $\gamma p\to K^0\pi^0\Sigma^+$ reaction was dominated by the mechanisms shown in Fig. \ref{fig:tree_level_ks}
where the $\Delta^*(1700)$ is excited by the photon and decays into $K^0\Sigma^{*+}$ and the $\Sigma^{*+}$ subsequently decays into $\pi^0\Sigma^+$. Since the cross section for this reaction is proportional to $(g_{\Delta^*K\Sigma^*})^2$, the agreement of the predicted cross section with experiment would provide support for the coupling provided by the theory with the assumption of the $\Delta^*(1700)$ as a dynamically generated resonance. The experiment for this reaction has been performed at ELSA and is presently under analysis. Preliminary results presented in the NSTAR05 workshop \cite{nanovanstar} agree with the theoretical predictions.

We would like to stress the fact that the couplings predicted by the theory based on the dynamical nature of the \dell resonance are by no means trivial. Indeed, should we assume that the \dell belongs to an $SU(3)$ decuplet as suggested in the PDG (table 14.5) \cite{Eidelman:2004wy} it is easy to see that the couplings to the $\Delta\pi$, $\Sigma^* K$, $\Delta\eta$ states in $I=3/2$ are proportional to $\sqrt{5/8}$, $\sqrt{1/4}$, $\sqrt{1/8}$, respectively.  The squares of these coefficients are proportional to $1$, $2/5$, $1/5$, respectively, compared to the squares of the coefficients of the dynamically generated resonance, $1$, $11.56$, $4.84$. It was noted in \cite{decu_ss} that the strength of the $\Delta\pi$ coupling of the dynamically generated model was consistent with the experimental branching ratio of the \dell. Hence, this means that the dynamically generated model produces considerable strength for the $\Sigma^*K$ and $\Delta\eta$ channels in absolute terms. With respect to the decuplet assumption of the PDG one obtains factors $27.5$ and $24$ larger for the square of the couplings to $\Sigma^*K$ and $\Delta\eta$, respectively. These large couplings indicate that, even if the \dell resonance is somewhat sub-threshold for the $\pi N\to\Sigma^*K$ and $\pi N\to\Delta\eta$ reactions, the combination of these couplings and the \dell width ($\sim 300$ MeV) should make this resonance play an important role in those reactions close to their thresholds.

It is clear that ultimately it is the consistency of the predictions of the theoretical models that builds up support for the theory. Hence, it is straightforward to suggest an additional reaction with a similar mechanism as in Fig. \ref{fig:tree_level_ks} but rather with the $\Sigma^{*}\to\pi\Lambda$ decay. Since the branching ratio for $\pi\Lambda$ decay of the $\Sigma^*$ is 88\%, the cross section would be reasonably larger than for $\gamma p\to K^0\pi^0\Sigma^+$ and one would have extra tests for the model. 

Additional tests can be also done with the related reactions, $\pi^- p\to K^0\pi^0\Lambda$ and all the other pion-induced reactions mentioned in the abstract, for which some data on cross sections are already available \cite{Thomas:1973uh,Hanson:1972zz,Grether:1973sz,dahl,curtis}. It is quite interesting to recall that in the theoretical model studied in \cite{Hyodo:2003jw}, which was based on the excitation of the $\Lambda(1405)$, the $K^0K^-p$, $K^0\overline{K}^0n$, $K^0\pi^+\Sigma^-$, and $K^0\pi^-\Sigma^+$ channels were reproduced within 25\%, while the cross section for the $K^0\pi^0\Lambda$ channel predicted was $6\;\mu$b compared to the $104\pm 8\;\mu$b of the experiment. The lack of the $\Sigma^*(1385)$ resonance in the model of \cite{Hyodo:2003jw}, which relied upon the final state interaction of the particles to generate dynamically the resonances, did not allow one to make a realistic approach for the $K^0\pi^0\Lambda$ final state, which in \cite{Thomas:1973uh} was shown to be dominated by $K^0\Sigma^{*0}$ production. The $\Sigma^*(1385)$, as all the other elements of the decuplet of the $\Delta(1232)$, does not qualify as a dynamically generated resonance, and is indeed a building block to generate other resonances like the $\Lambda^*(1520)$ or $\Delta^*(1700)$.

While at the time of \cite{Hyodo:2003jw} the information on the $\Delta^*(1700)\to K\Sigma^*$ coupling was not available, the works of \cite{lutz,decu_ss}, and particularly \cite{decu_ss} where the coupling is evaluated, have opened the door to tackle this reaction and this is one of the aims of the present work.

The data for the $\pi^-p\to K^0\pi^0\Lambda$ reaction from \cite{Thomas:1973uh} is at $\sqrt{s}=2020$ MeV which is about 320 MeV above the \dell peak, potentially too far away to claim dominance of the $\Delta^*(1700)$, but there are also data at lower energies \cite{dahl,curtis} around $\sqrt{s}=1930$-1980 MeV. On the other hand there are data \cite{Hanson:1972zz,Grether:1973sz} for other reactions, $\pi^+ p\to K^+\pi^+\Lambda$, $\pi^+ p\to K^+\pi^+\Sigma^0$, $\pi^+ p\to K^+\pi^0\Sigma^+$, and $\pi^+ p\to \eta\pi^+ p$ at energies around $\sqrt{s}=1800$ MeV which allow us to make a more direct comparison with the theoretical predictions. In the next section we study these reactions. 
Similarly, there are also some data for the $\gamma p\to K^0\pi^+\Lambda$, $K^+\pi^-\Sigma^+$, $K^+\pi^+\Sigma^-$ reactions \cite{Erbe:1970cq,Erbe:2,cmc} and we shall also address them in the same context.

\section{The model for the $\pi p\to K\pi\Lambda$, $K\pi\Sigma$, $K\bar{K}N$, $\eta\pi N$ reactions}
\begin{figure}
\includegraphics[width=12cm]{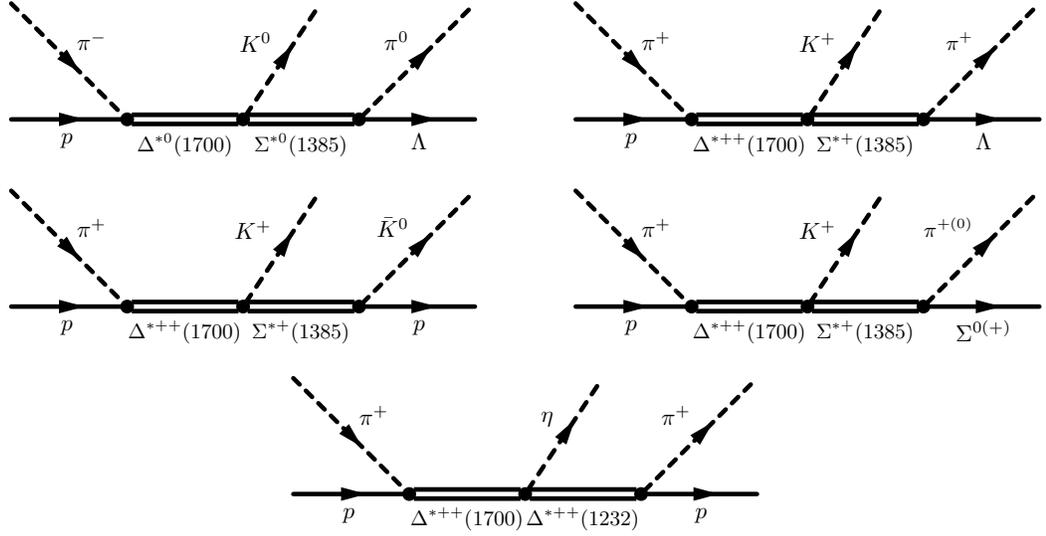}
\caption{Tree level contributions for the pion-induced strangeness production via the \dell.}
\label{fig:tree_level_new}
\end{figure}

In the work of \cite{Doring:2005bx} on the $\gamma p\to K^0\pi^0\Sigma^+$ reaction, the diagram of Fig. \ref{fig:tree_level_ks} was evaluated together with many loop diagrams involving rescattering of the $K\Sigma$ state which are the building blocks together with $\pi N$ and others of the $N^*(1535)$ resonance. It was found there that the tree level diagram was dominant. For energies of $\sqrt{s}=2000$ MeV, hence 300 MeV above the $\Delta^*(1700)$ nominal mass, the loop terms could provide a contribution of about 30\% with large uncertainties. On the other hand the results obtained for the cross section find support in preliminary experimental results presented at the NSTAR05 workshop \cite{nanovanstar}. With this information at hand we have good justification to propose that the dominant mechanism for the $\pi^-p\to K^0\pi^0\Lambda$, $\pi^+ p\to K^+\pi^+\Lambda$, $K^+\bar{K}^0p$, $K^+\pi^+\Sigma^0$, $K^+\pi^0\Sigma^+$, and $\eta\pi^+ p$ reactions is given by the diagrams of Fig. \ref{fig:tree_level_new}.

All the elements of these diagrams are at hand from the works of \cite{decu_ss,Doring:2005bx}. The new information needed here is the $\pi N$ coupling to the $\Delta^*(1700)$ which is not an ingredient of the building blocks in the studies of \cite{lutz,decu_ss} that only take into account the interaction of the octet of pseudoscalar mesons with the decuplet of baryons. Thus, we take this information from experiment by looking at the branching ratio in the PDG \cite{Eidelman:2004wy}. In spite of the larger phase space for decay into this channel the branching ratio to $\pi N$ is only 10-20\%. 

By taking into account that the coupling of $\pi N$ to $\Delta^*(1700)(3/2^-)$ is in $d$-wave, the structure of the $\Delta^*\pi N$ vertex is most conveniently written as
\be
-it_{\Delta^*(1700)\to\pi N}&=&-ig^{(d)}_{\pi N\Delta^*}\;{\cal C}(1/2\;2\;3/2;m,M-m)\;Y^*_{2,\;m-M}(\hat{\bf k})(-1)^{M-m}\sqrt{4\pi}	\label{d-wave-vertex}
\ee
as in Ref. \cite{Sarkar:2005ap} (see their Eq. (25)) to account for the $\Lambda^*(1520)\to\pi\Sigma$ coupling. In Eq. (\ref{d-wave-vertex}) the Clebsch Gordan coefficient accounts for the matrix element of the rank two spin operator needed to couple the spherical harmonic $Y_2$ to a scalar. The quantities $M,m$ are the third components of the spin of the $\Delta^*$ and the nucleon and ${\bf k}$ is the momentum of the pion. The $\pi N$ state is in $I=3/2$. 

With the parametrization of Eq. (\ref{d-wave-vertex}) the partial decay width of the $\Delta^*$ to $\pi N$ is written as
\be
\Gamma_{\Delta^*\to\pi N}=\frac{(g^{(d)}_{\pi N\Delta^*})^2}{2\pi}\;\frac{M_N}{M_{\Delta^*}}\;k_\pi.
\ee
By taking the width of the $\Delta^*$, $\Gamma_{\Delta^*}=300\pm 100$ MeV, and the $\pi N$ branching ratio of $15\pm 5$\% and summing the errors in quadrature we obtain the value 
\be
g^{(d)}_{\pi N\Delta^*}=0.94\pm 0.20
\label{gdwave}
\ee
which will necessarily lead to uncertainties of the order of 50\% in the cross section.
Given the isospin decomposition of the $\pi^-p$ state in $I=1/2,\;3/2$, the coupling $g_{\pi N\Delta^*}$ (in $I=3/2$) has to be multiplied by $\sqrt{1/3}$ to account for the coupling of the $\pi^-p$ state to the $\Delta^*$ and by $-1$ to account for the coupling of $\pi^+p$ to $\Delta^{*++}$ (although irrelevant for the cross section we use the isospin phase convention $|\pi^+\rangle=-|1,1\rangle$). This means
\be
g^{(d)}_{\pi^- p\Delta^{*0}}=\sqrt{\frac{1}{3}}\,g^{(d)}_{\pi N\Delta^*}, \; g^{(d)}_{\pi^+ p\Delta^{*++}}=-g^{(d)}_{\pi N\Delta^*}.
\label{gd}
\ee

There is another point worth noting which is that due to the $d$-wave character of the $\pi N\Delta^*$ vertex, the coupling $g_{\pi N\Delta^*}$ implicitly incorporates $k_\pi^2$ for the on-shell value of the pion momentum in the $\Delta^*\to\pi N$ decay. When we extrapolate beyond the resonance energy, as will be the case here, we must then use
\be
g_{\pi N\Delta^*}\to g_{\pi N\Delta^*}({\rm on\;shell})\;\frac{BW(k_\pi\; R)}{BW(k^{{\rm on}}_\pi\; R)}
\label{gchange}
\ee
where $BW(\cdot)$ is the Blatt and Weisskopff penetration factor \cite{blatt,Manley:1984jz,Manley:1992yb}
\be
BW(x)=\frac{x^2}{\left(9+3x^2+x^4\right)^{1/2}}
\ee
and $R=0.4$ fm according to best fits of \cite{Manley:1984jz}. 

The other couplings needed are those of the decuplet to the meson and baryon octets.
For the vertices with $\Sigma^*(1385)$ or $\Delta(1232)$ decay in the diagrams of Fig. \ref{fig:tree_level_new} we use the chiral Lagrangian \cite{Butler:1992pn}
\be
{\cal L}=
{\cal C}\left(\sum_{a,b,c,d,e}^{1,\cdots,3}\epsilon_{abc}\;\overline{T}^{ade} \;\overline{u}_\mu\;A_{d}^{b,\mu} \;B^c_e
+\sum_{a,b,c,d,e}^{1,\cdots,3}\epsilon^{abc}\;{\overline B}^e_c \;A^{d}_{b,\mu} \;T_{ade}\;u^\mu\right).
\label{precise_L}
\ee
The Rarita-Schwinger fields $u$ are defined as in \cite{decu_ss} and the flavor tensor $T$ is given in \cite{Butler:1992pn}.
In Eq. (\ref{precise_L}), the quantity $A^\mu$, which is proportional to the axial current, is expanded up to one meson field, 
\be
A^\mu=\frac{i}{2}\left(\xi\partial^\mu\xi^\dagger-\xi^\dagger\partial^\mu\xi\right)\stackrel{{\rm one}\;\Phi}{\longrightarrow} 
\frac{\partial^\mu\Phi}{\sqrt{2}f_\pi}, \quad
\xi=\exp\left(\frac{i\Phi}{\sqrt{2}f_\pi}\right),
\ee
$\Phi, B,\overline{B}$ are the standard $SU(3)$ matrices of the meson and baryon fields, and $f_\pi=93$ MeV.
The Lagrangian in Eq. (\ref{precise_L}) allows one to relate the $\pi\Lambda\Sigma^*$ coupling to $\pi N\Delta$. For the vertex one finds
$-it_{B^*\to B\Phi}=a\,{\bf S}\cdot {\bf q}$ where
\begin{align}
a_{\Sigma^{*0}\to\pi^0\Lambda} &=\frac{0.82}{\sqrt{2}}\,\frac{f_{\pi N\Delta}}{m_\pi},    & 
a_{\Sigma^{*+}\to\pi^+\Lambda} &=-\,\frac{0.82}{\sqrt{2}}\,\frac{f_{\pi N\Delta}}{m_\pi}, & 
a_{\Sigma^{*+}\to\bar{K}^0 p}  &=\frac{2\sqrt{6}}{5}\,\frac{D+F}{2f_\pi},\non
a_{\Sigma^{*+}\to\pi^+\Sigma^0}&=-\,\frac{0.78}{\sqrt{6}}\,\frac{f_{\pi N\Delta}}{m_\pi}, & 
a_{\Sigma^{*+}\to\pi^0\Sigma^+}&=\frac{0.78}{\sqrt{6}}\,\frac{f_{\pi N\Delta}}{m_\pi},    &
a_{\Delta^{++}\to\pi^+p}       &=\frac{f_{\pi N\Delta}}{m_\pi} 
\label{itsigstar}
\end{align}
with ${\bf q}$ the momentum of the outgoing meson in the $\Sigma^*$ or $\Delta$ rest frame and ${\bf S}$ the spin transition operator from $3/2$ to $1/2$ normalized as
\be
\langle M|S^\dagger_\mu |m\rangle={\cal C}\left(1/2\;1\;3/2;m\;\mu\;M\right).
\ee
For the $f_{\pi N\Delta}$ we take $f_{\pi N\Delta}=2.13$ to give the experimental $\Delta$ width.  
Note that $SU(3)$ symmetry, implicit in Eq. (\ref{itsigstar}), is not exact. In order to obtain the experimental $\Sigma^*\to\pi\Lambda$, $\Sigma^*\to\pi\Sigma$ widths one can fit the coupling ${\cal C}$ from Eq. (\ref{precise_L}) to the branching ratios from the PDG \cite{Eidelman:2004wy}. This leads to a correction which appears as a numerical factor in Eq. (\ref{itsigstar}). For the $\Sigma^*\to\bar{K}N$ decay which is physically closed we use a $SU(6)$ quark model prediction \cite{Oset:2000eg}.

The couplings from \cite{decu_ss} of the \dell to its $s$-wave decay channels are given for $I=3/2$ and counting the isospin decomposition of $K^0\Sigma^{*0}$, $K^+\Sigma^{*+}$, $\eta\Delta$, we find the couplings
\be
g_{K^0\Sigma^{*0}\Delta^{*0}}=\sqrt{\frac{2}{3}}\;g_{K\Sigma^*\Delta^*}, \quad g_{K^+\Sigma^{*+}\Delta^{*++}} =g_{K\Sigma^*\Delta^*},\quad g_{\eta\Delta^{++}\Delta^{*++}}=g_{\eta\Delta\Delta^{*}}.
\label{gstar}
\ee
Altogether, our amplitudes for the diagrams of Fig. \ref{fig:tree_level_new} become
\be
-it&=&a\;{\bf S}\cdot {\bf q}\;G\;\frac{1}{\sqrt{s_{\Delta^*}}-M_{\Delta^*}+\frac{i\Gamma_{\Delta^*}(s_{\Delta^*})}{2}}\;g_j\;g^{(d)}_i\;\frac{BW(k R)}{BW(k^{\rm on}R)}\non &&\non
&\times&
{\cal C}\left(1/2\;2\;3/2;m,M-m\right)Y_{2,\;m-M}(\hat{{\bf k}}) (-1)^{M-m}\sqrt{4\pi}
\label{tfirst}
\ee
with $k^{\rm on}$ the pion momentum in the $\pi N$ decay of the $\Delta^*(1700)$ at rest. Depending on the process, $G=1/(\sqrt{s_{B^*}}-M_{B^*}+i/2\,\Gamma_{B^*}(\sqrt{s_{B^*}}))$ is the $\Sigma^*(1385)$ or $\Delta(1232)$ propagator; $a$ in Eq. (\ref{tfirst}) is given by Eq. (\ref{itsigstar}) and $g_j, \;g^{(d)}_i$ by Eqs. (\ref{gstar}) and (\ref{gd}), respectively.

For the momentum-dependent width of the $\Sigma^*(1385)$, we have taken into account the $p$-wave decays into $\pi\Lambda$ and $\pi\Sigma$ with their respective branching ratios of 88\% and 12\%. For the width of the $\Delta^*(1700)$ we have included the dynamics of the decay into $\Delta^*\to N\rho (N\pi\pi)$ and $\Delta^*\to \Delta\pi (N\pi\pi)$ in the same way as in Ref. \cite{Doring:2005bx}. 
We introduce a novelty with respect to Ref. \cite{Doring:2005bx} where the $\rho N$ decay of the $N^*(1520)$, \dell was considered in $s$-wave. In the case of the \dell the $d$-wave $\rho N$ decay is mentioned in the PDG \cite{Eidelman:2004wy} as existing but with an undetermined strength. We have adopted here to take a $\rho N$ strength in $s$-wave of $37$\%, and $5$\% for $\rho N$ in $d$-wave after a fine tuning to the data. 
For energies close to the \dell only the total width matters and this is about 300 MeV in our case.
For the width of the $\Delta^*\to\rho N$ in $d$-wave we use the formula of Ref. \cite{Doring:2005bx} (the equation before Eq. (37)) by multiplying the numerator by $BW(|{\bf q}_1-{\bf q}_2|^2)$ and one coupling is adjusted to get the 5\% of the branching ratio used.

We should mention here that having $\rho N$ decay in $s$-wave or $d$-wave produces large differences at energies around $\sqrt{s}=2$ GeV and beyond, but only moderate differences close to the \dell peak or 100-150 MeV above it.
 
For the \dell width from decay into $\pi N$ in $d$-wave, an additional Blatt-Weisskopff factor is applied to be consistent with Eq. (\ref{gchange}). With the partial width $\Gamma^0_{\pi N}=0.15\times 300$ MeV, the width is given by \cite{Manley:1992yb}
\be
\Gamma_{\pi N}&=&\Gamma^0_{\pi N}\;\frac{k\;BW^2(k R)\;M_{\Delta^*}}{k^{\rm on}\;BW^2(k^{\rm on} R)\;\sqrt{s}}
\label{momentum_width}
\ee
and the total width is the sum of the partial widths of the decay modes. The same is done for the $d$-wave of the $\rho N$ decay.

Summing $|t|^2$ from Eq. (\ref{tfirst}) over the final states, the sum does not depend on the original proton polarization. We are free to choose $m=1/2$ in which case the amplitude of Eq. (\ref{tfirst}) becomes
\be
-it&=&\frac{a}{\sqrt{3}}\;G\;\frac{1}{\sqrt{s_{\Delta^*}}-M_{\Delta^*}+\frac{i\Gamma_{\Delta^*}(s_{\Delta^*})}{2}}\;
g_j\;g^{(d)}_i\;
\frac{BW(k_\pi R)}{BW(k_\pi^{\rm on}R)}\;
\Bigg\{\begin{array}{lll}
&2q_z&;m'=+1/2\\ \vspace*{-0.2cm} & &\\
-&\left(q_x+iq_y\right)&;m'=-1/2.
\end{array}
\label{tsecond}
\ee
In the sum of $|t|^2$ over the final states of $\Lambda$ with $m'=1/2,-1/2$ the part corresponding to the curly bracket in Eq. (\ref{tsecond}) will become
\be
4q_z^2+q_x^2+q_y^2=3q_z^2+{\bf q}^2
\label{qxqyqz}
\ee
which gives an angular distribution proportional to $(3\cos^2\theta+1)$ in the angle of the outgoing meson from the $\Sigma^*$ or $\Delta$ decay with respect to the initial $\pi^-$ direction for this initial proton polarization. When integrating over angles Eq. (\ref{qxqyqz}) can be replaced by $2{\bf q}^2$. For the first reaction from Fig. \ref{fig:tree_level_new} the $(\pi^0\Lambda)$ invariant mass distribution is then given by
\be
\frac{d\sigma}{dM_I(\pi^0\Lambda)}=\frac{M_p M_\Lambda}{\lambda^{1/2}(s,m_\pi^2,M_p^2)}\;\frac{q_{\pi^0}q_{K^0}}{(2\pi)^3\sqrt{s}}\overline{\sum}\sum |t|^2
\label{dsdmi}
\ee
in terms of the ordinary K\"allen function $\lambda^{1/2}$ and $t$ from  Eq. (\ref{tsecond}), $q_{K^0}=\lambda^{1/2}(s,M_I^2,m_{K^0}^2)/(2\sqrt{s})$,
$q\equiv q_{\pi^0}=\lambda^{1/2}(M_I^2,m_{\pi^0}^2,M_\Lambda^2)/(2M_I)$. Furthermore, the variables in Eq. (\ref{tsecond}) take the values $\sqrt{s_{\Delta^*}}=\sqrt{s}$, $\sqrt{s_{\Sigma^*}}=M_I$ and the total cross section is given by integrating over $M_I$ in Eq. (\ref{dsdmi}).
The generalization to other channels is straightforward by changing the masses and corresponding momenta.

\section{The model for the $\gamma p\to K\pi\Lambda$, $k\pi\Sigma$, $\eta\pi p$ reactions}
The photon coupling to the \dell resonance is taken from \cite{Nacher:2000eq,Doring:2005bx}. The processes $\gamma p\to K^0\pi^+\Lambda$, $K^+\pi^-\Sigma^+$, $K^+\pi^+\Sigma^-$, $K^0\pi^0\Sigma^+$, $\eta\pi^0 p$ are given by diagrams similar to those of Fig. \ref{fig:tree_level_new} with the incoming pion replaced by the photon. In view of this it is very easy to modify the pion-induced amplitudes of the previous section to write the photon-induced ones. 
The contributions for the first four reactions are given by
\be
  T_{\gamma p\to MMB}^{\rm 
BG}&=&b\;\frac{2}{5}\frac{D+F}{2f_\pi}\;g_{K\Sigma^*\Delta^*}
\;G_{\Sigma^*}(\sqrt{s_{\Sigma^*}})\; G_{\Delta^*}(\sqrt{s_{\Delta^*}})\vec{S}\cdot {\bf
p}_\pi\non
&\times &
\left[-ig_1'\frac{\vec{S}^\dagger\cdot {\bf 
k}}{2M}(\vec{\sigma}\times {\bf
k})\cdot\vec{\epsilon}-\vec{S}^\dagger\cdot\vec{\epsilon}\left(g_1'(k^0+\frac{{\bf
k}^2}{2M})+g_2'\sqrt{s}\;k^0\right)\right]\non
\label{tree_level_kzero_sigplus}
\ee
with
\be
b_{\gamma p\to K^0\pi^+\Lambda}=1.04\sqrt{3}, \quad b_{\gamma p\to K^+\pi^-\Sigma^+}=-\sqrt{2},\quad b_{\gamma p\to K^+\pi^+\Sigma^-}=\sqrt{2},
\quad b_{\gamma p\to K^0\pi^0\Sigma^+}=1
\ee
which follows from Eq. (49) from Ref. \cite{Doring:2005bx} by applying the corresponding changes in isospin factors and corrections due to $SU(3)$ breaking in the $\Sigma^*(1385)$ decay similar as in Eq. (\ref{itsigstar}). Cross sections and invariant masses for the photon reactions are given by Eq. (\ref{dsdmi}) with the corresponding changes in masses (e.g., $m_\pi\to 0$ in $\lambda^{1/2}$). For the $\gamma p\to \eta\pi^0 p$ reaction we replace the factor before the brackets in Eq. (\ref{tree_level_kzero_sigplus}) by
\be
-\sqrt{\frac{2}{3}}\,g_{\eta\Delta\Delta^*}\,\frac{f_{\pi N\Delta}}{m_\pi}\, G_{\Delta^*}(\sqrt{s_{\Delta^*}})\,G_{\Delta}(\sqrt{s_{\Delta}}).
\ee
The $\gamma p\to\eta\pi^0$ and $\gamma p\to K^0\pi^0\Sigma^+$ reactions have been derived in Ref. \cite{Doring:2005bx}. Besides the tree level amplitudes there are one-loop transitions between \dell and another dynamically generated resonance, the $N^*(1535)$. The latter terms are not important at the high energies in which we are currently interested because the $N^*(1535)$ is off-shell.

\section{Results and discussion}
\begin{figure}
\includegraphics[width=14cm]{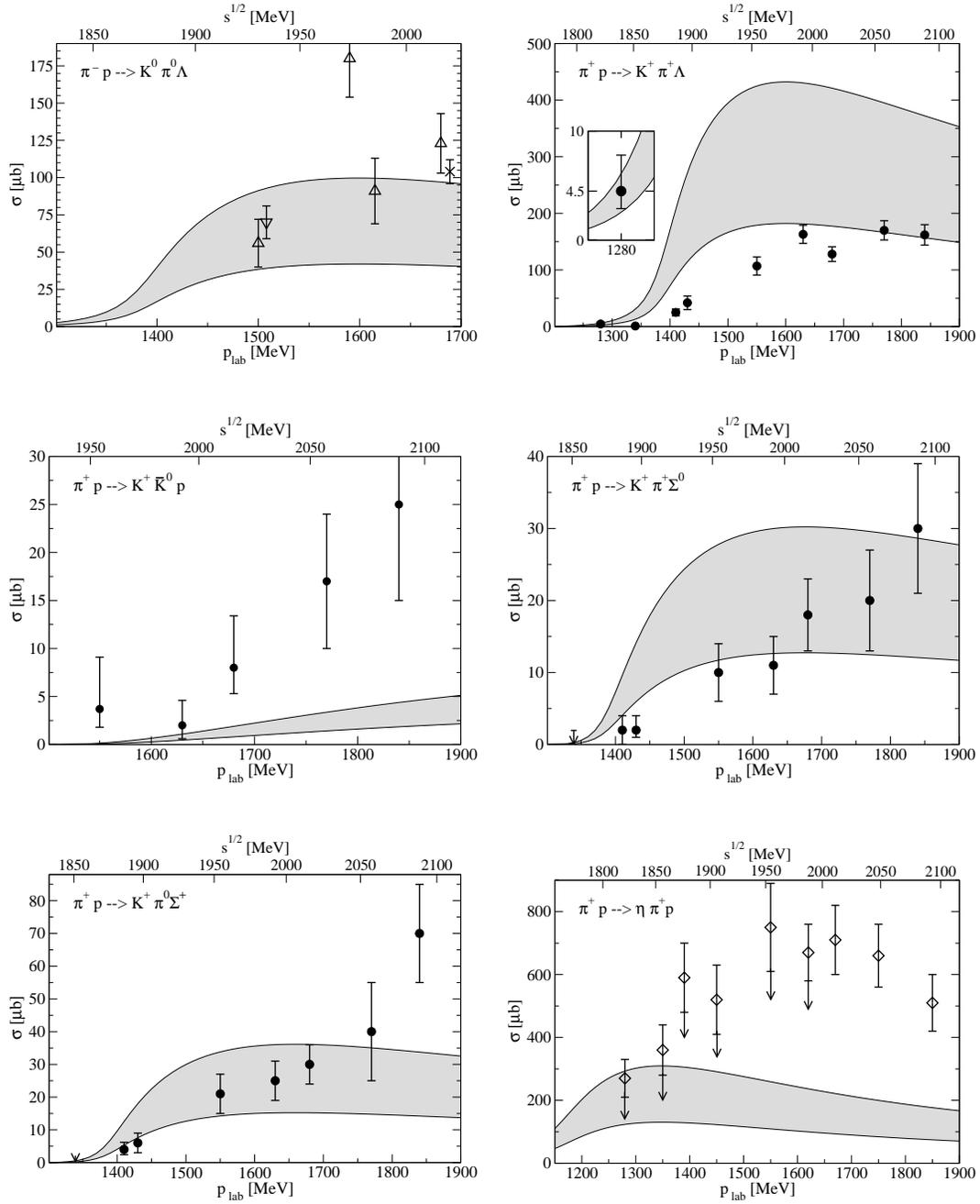}
\caption{Total cross sections for the pion-induced reactions. Data are from 
\cite{dahl} (triangles up),
\cite{curtis} (triangle down),
\cite{Thomas:1973uh} (cross),
\cite{Hanson:1972zz} (dots), 
\cite{Grether:1973sz} (diamonds).
For the latter data, it is indicated that these are upper limits below $p_{\rm}=1.67$ GeV as in Ref. \cite{Grether:1973sz}.}
\label{fig:sigma}
\end{figure}
\begin{figure}
\includegraphics[width=17cm]{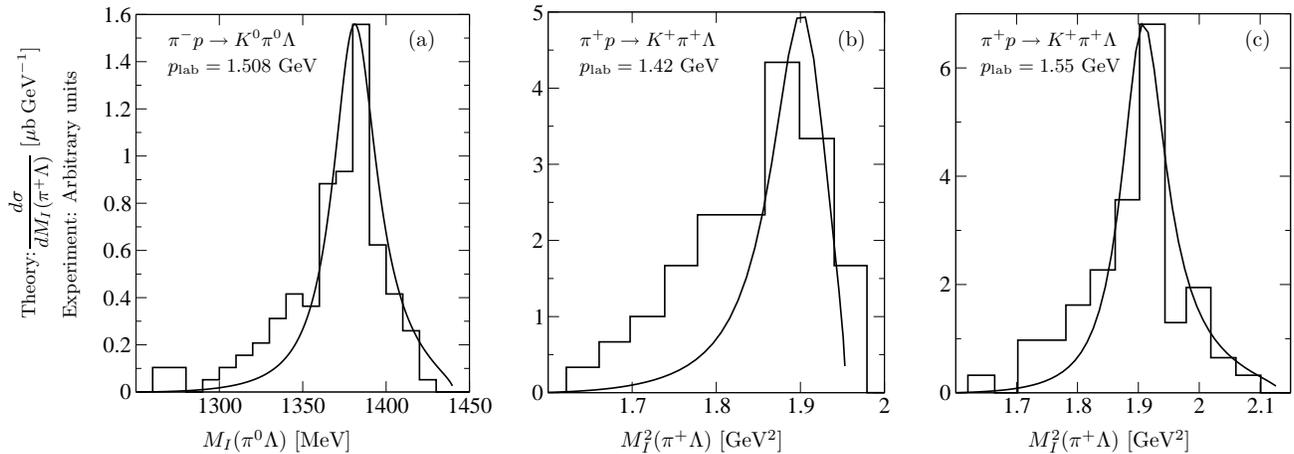}
\caption{Invariant mass spectra for the $\pi^-p\to K^0\pi^0\Lambda$ and $\pi^+ p \to K^+ \pi^+ \Lambda$ reactions. Experimental distributions (arbitrary units) are from \cite{curtis} and \cite{Hanson:1972zz}, respectively.}
\label{fig:dsdmi}
\end{figure}
\begin{figure}
\includegraphics[width=8cm]{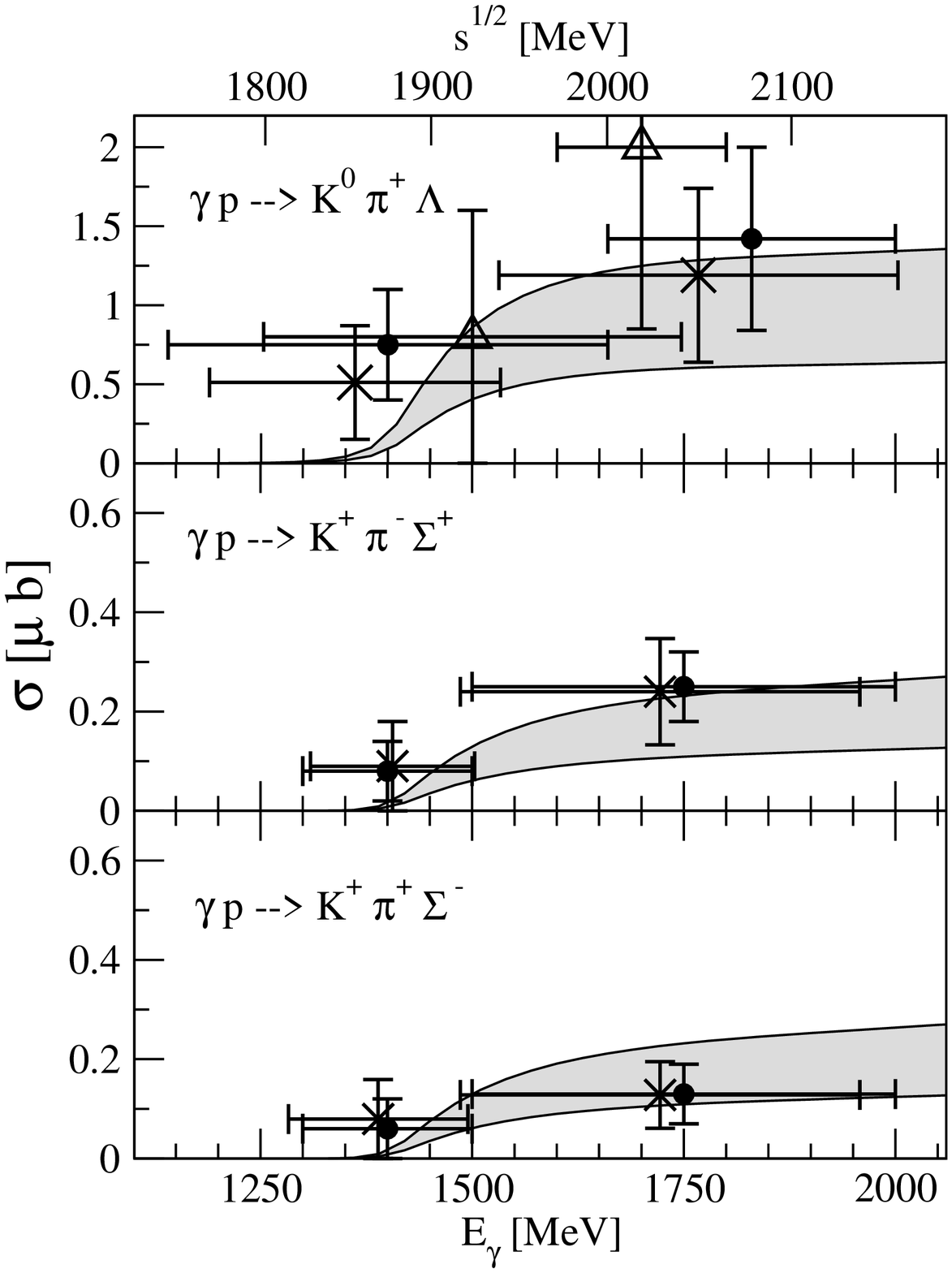}

\vspace*{0.5cm}

\includegraphics[width=7cm]{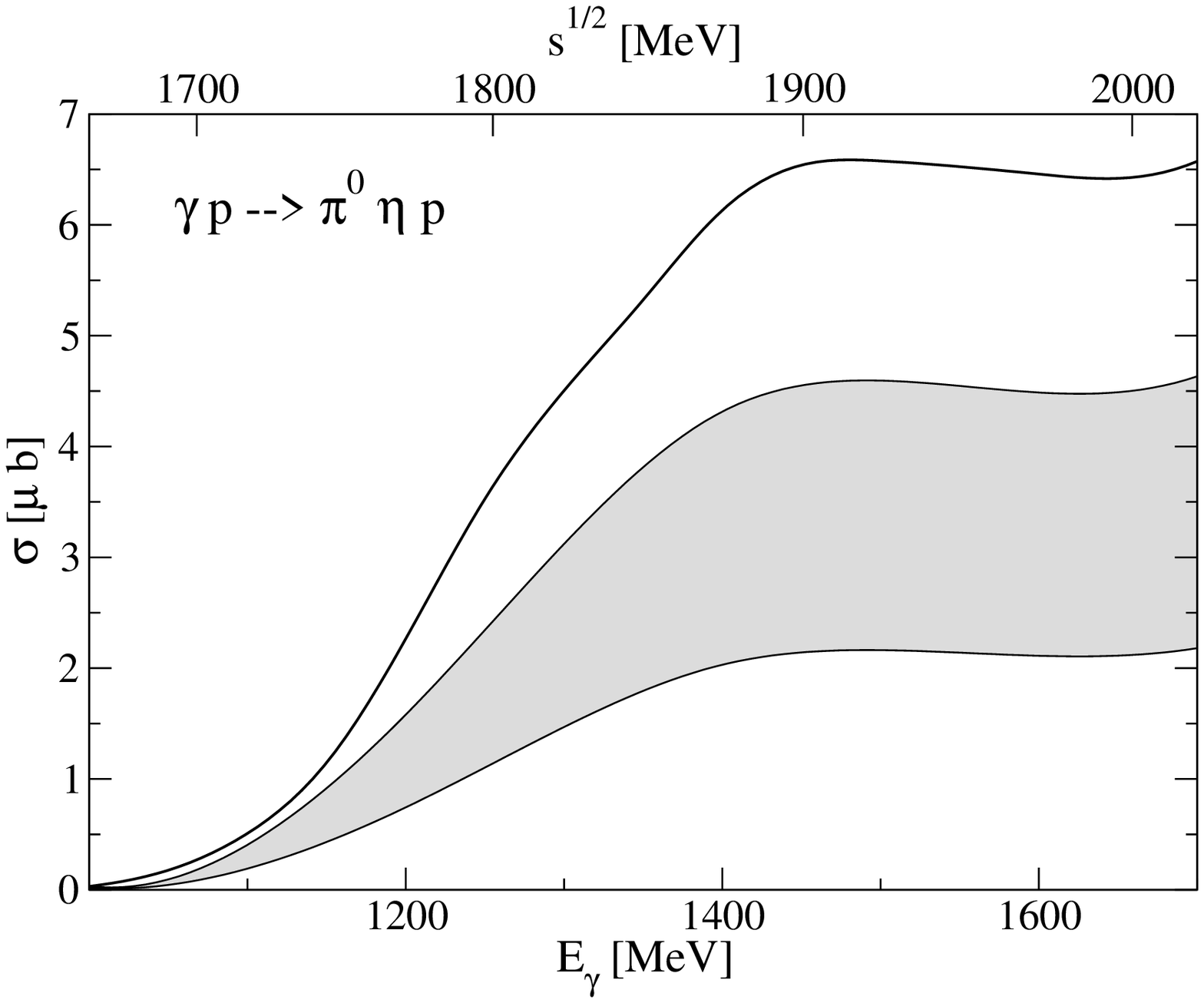}
\hspace*{0.5cm}
\includegraphics[width=7.5cm]{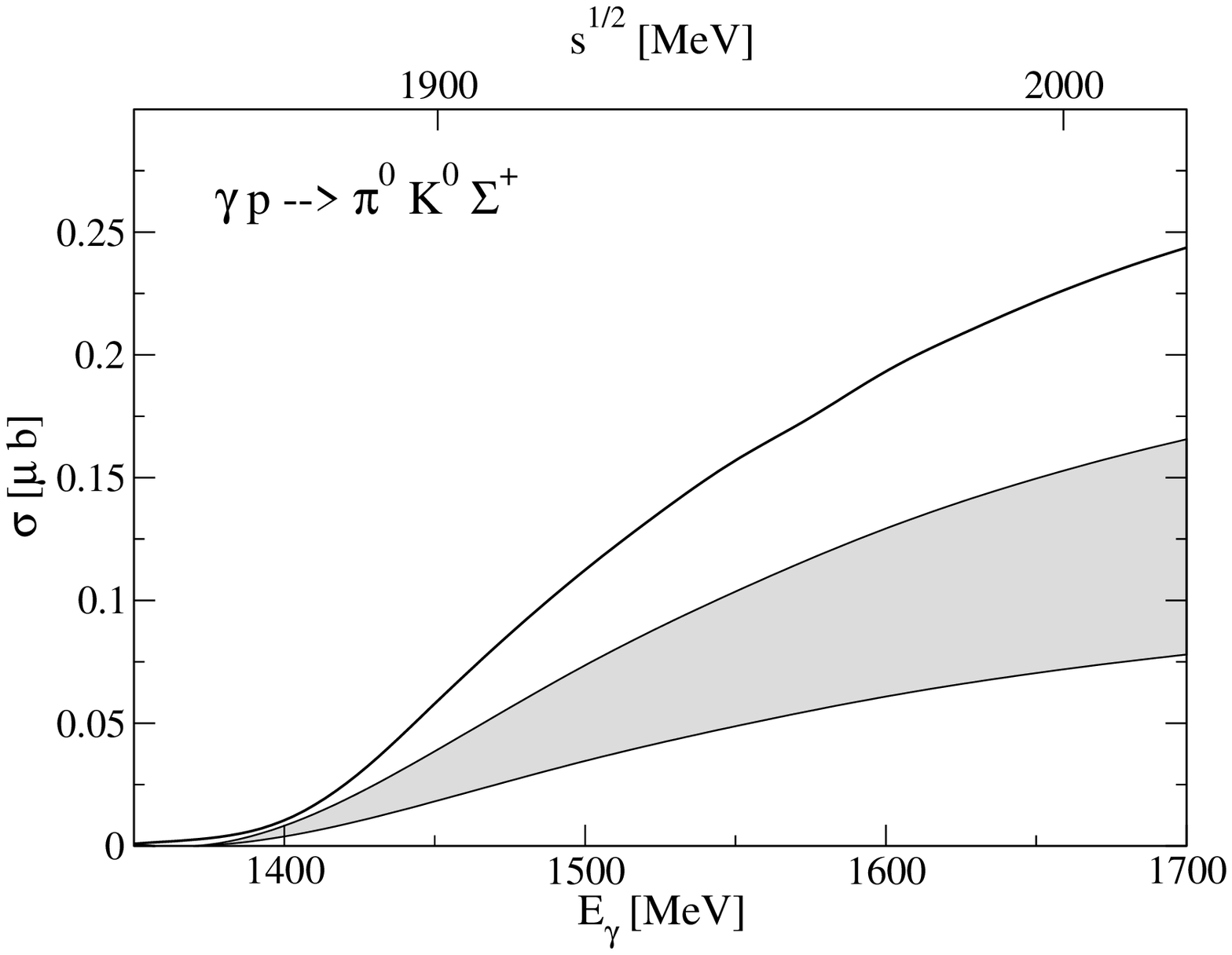}
\caption{Photoproduction of strange and $\eta$ particles. Data are from 
\cite{Erbe:1970cq} (dots), \cite{Erbe:2} (crosses), and \cite{cmc} (triangles up). The latter data are the sum of $K\Lambda\pi$ and $K\Sigma\pi$ final states. 
The two lower plots show the update of the predictions from Ref. \cite{Doring:2005bx} (gray bands) compared to the results from \cite{Doring:2005bx} (solid lines).}
\label{fig:photo}
\end{figure}

In Figs. \ref{fig:sigma} to \ref{fig:photo} we show the results for the pion- and photon-induced reactions.
In Fig. \ref{fig:sigma} we show the cross sections for the $\pi^- p \to K^0 \pi^0 \Lambda$, 
$\pi^+ p \to K^+ \pi^+ \Lambda$,  $\pi^+ p \to K^+ \bar{K}^0 p$,
$\pi^+ p \to K^+ \pi^+ \Sigma^0$, $\pi^+ p \to K^+ \pi^0 \Sigma^+$, and 
$\pi^+ p \to \eta \pi^+ p$ reactions.  The theoretical results are plotted in
terms of a band. This band corresponds to taking the $\pi N \Delta^*$ coupling
with its uncertainties (from the experimental branching ratio ) quoted  in Eq. 
 (\ref{gdwave}).  
 
Much of the data in Fig. \ref{fig:sigma} are for energies $\sqrt{s}>1950$ MeV, which, even taking into account the 300 MeV width of the $\Delta^*(1700)$, are relatively far away from the $\Delta^*(1700) $ peak.
   In a situation like this it is logical to assume that other channels not
   related to the $\Delta^*(1700)$ could also play a role.  Indeed, other partial
   waves in  $\pi N$ scattering are equally important in this energy region 
  \cite{Arndt:2003if,Vrana:1999nt}. However, there are reasons to assume that they do not
  couple strongly to the $K \pi \Lambda, \,K\pi\Sigma$ in the final state at the lower energies, as we shall discuss at the end of this section. In any case, even at the higher energies, the order of magnitude
  predicted for the cross section is correct.  

For the $\pi^-p\to K^0\pi^0\Lambda$ reaction we have data around $\sqrt{s}=1930$ MeV and the theory agrees with those data. Furthermore, as we can see in \cite{curtis}, the $\pi^0\Lambda$ mass spectrum is totally dominated by the $\Sigma^*(1385)$ and, as shown in Fig. \ref{fig:dsdmi} (a),
the theoretical predictions agree with these data. It is also very instructive to see that the angular distribution of the $\Sigma^*$ is practically flat \cite{curtis,goussu}, as our model predicts, given the $s$-wave coupling of the $\Delta^*(1700)$ to $K\Sigma^*$.

 Next we discuss the cross section
 for the $\pi^+ p \to K^+ \pi^+ \Lambda$ reaction.  The range of energies extends 
 now from about $\sqrt{s}=1800$ MeV on.  We can see that the order of magnitude
 of the cross sections from the theoretical band is correct, although the
 theoretical prediction is more than a factor of two bigger than data at around $\sqrt{s}=1900$ MeV. Yet, this
 apparently large difference should be viewed in perspective, which is
 provided by the cross section of the 
  $\pi^+ p \to K^+ \pi^+ \Sigma^0$ and  $\pi^+ p \to K^+ \pi^0 \Sigma^+$
 reactions. Indeed, given the larger coupling of the $\Sigma^*(1385)$ resonance to
 $\pi \Lambda$, with a branching ratio to this latter channel about one order of
 magnitude larger than for $\pi \Sigma$, the mechanism that we have should
 provide a cross section for the $\pi^+ p \to K^+ \pi^+ \Lambda$ reaction about
 one order of magnitude bigger than for the 
  $\pi^+ p \to K^+ \pi^+ \Sigma^0$ or  $\pi^+ p \to K^+ \pi^0 \Sigma^+$
  reactions. This is indeed the case, both in the theory and in the experiment. 
  We can see that in the region of energies below $\sqrt{s}< 1900$ MeV the
  agreement of the theory with the data is fine at the qualitative level for
  these two latter reactions. We can also see that the first data point for the $\pi^+p\to K^+\pi^+\Lambda$ reaction is in agreement with the theoretical prediction (see the insert in Fig. \ref{fig:sigma})
  
  The invariant mass spectra for different energies in the $\pi^+ p \to K^+ \pi^+ \Lambda$ reaction are shown in Fig. \ref{fig:dsdmi} (b), (c).
 These curves can be directly compared to the data of Ref. \cite{Hanson:1972zz}; the $\Sigma^*(1385)$ dominance in both theory and experiment is apparent. Note that, as mentioned in \cite{curtis}, the excess of strength at the lower shoulder is partly a result of the finite experimental resolution \cite{curtis,Hanson:1972zz}.
  
    In Fig. \ref{fig:sigma} we also plot the cross section for the 
 $\pi^+ p \to K^+ \bar{K}^0 p$ reaction.  We can see that the predicted cross
 sections are quite low compared with experiment.  This should be expected since
 our mechanism is doubly suppressed there, first from having the $\Delta^*(1700)$
 off shell, and second from also having off shell the $\Sigma^*(1385)$ decaying
 into $\bar{K} N$. It is thus not surprising that our mechanism produces these
 small cross sections.  The $\Sigma^*(1385)$ is, however, not off shell for the
 $\pi \Sigma$ and $\pi \Lambda$ in the final state, and even at the low energies
 of the figure the mass distribution for these two particles is dominated by
 the $\Sigma^*(1385)$ in the theory, and this is also the case in the experiment
 as mentioned above.
 
   Finally, we also show in Fig. \ref{fig:sigma} the cross section for the 
$\pi^+ p \to \eta \pi^+ p$ reaction.  Here the mechanism is also $\Delta^*(1700)$
production but it decays into $\eta \Delta(1232)$ followed by 
$\Delta(1232) \to \pi N$. The agreement of the theory with the data is fair for
low energies, even more when we read the caution statement in the experimental
paper \cite{Grether:1973sz} warning that the data are overestimated below $p_{{\rm lab}}=1670$ MeV. Once again it is
worth noting that below $p_{{\rm lab}}=1670$ MeV the cross sections are a factor fifty 
larger than  for $\pi^+ p \to K^+ \pi^0\Sigma^+$ or 
$\pi^+ p \to K^+ \pi^+ \Sigma^0$. The theory is producing these large order of
magnitude changes in the cross sections correctly. 

  Next we discuss the photonuclear cross sections. In Fig. \ref{fig:photo}
   we can see the
  cross sections for the $\gamma p \to K^0 \pi^+ \Lambda$,  
$\gamma p \to K^+ \pi^- \Sigma^+$, $\gamma p \to K^+ \pi^+ \Sigma^-$,
$\gamma p \to K^0 \pi^0 \Sigma^+$ and $\gamma p \to \eta \pi^0 p$ reactions.  
For the first three reactions we have some data from \cite{Erbe:1970cq,Erbe:2,cmc}.  Only a few data
points appear in the region below $\sqrt{s} =1900$ MeV and, furthermore, the
data have large uncertainties, both in the magnitude of the cross section and in
the value of $\sqrt{s}$.  Nevertheless, the data are valuable in this global
analysis that we are doing. In the first place we can see that within errors, the
agreement of theory and experiment is fair. Yet, more significant is the ratio
of more than one order of magnitude, both in the theory and experiment for the cross
sections of the $\gamma p\to \eta\pi^0 p$ and the $\gamma p \to K^+ \pi^+ \Sigma^-$ or
$\gamma p \to K^+ \pi^- \Sigma^+$ reactions.
This follows the same trend as the $\pi^+ p \to \eta\pi^+ p$
and $\pi^+ p \to K^+ \pi^+ \Sigma^0$ or  $\pi^+ p \to K^+ \pi^0 \Sigma^+$
  reactions discussed above, supporting also the dominance of the same mechanisms
in the reactions.  
It is also worth noting that the dominance of the $\Sigma^*(1385)$ production for the $\gamma p\to K^0\pi^+\Lambda$ reaction is also mentioned in Ref. \cite{Erbe:2}.

In the figure we also plot the cross sections for the 
$\gamma p \to K^0 \pi^0 \Sigma^+$ and $\gamma p \to \eta \pi^0 p$ reactions
measured at ELSA and which are presently being analyzed \cite{nanovanstar}.  These
latter two reactions were studied in \cite{Doring:2005bx} with far more detail
including many more mechanisms.  Yet, it was this detailed study that showed the
dominance of the reaction mechanism considered in this paper.  We also found
there that around $\sqrt{s}= 2000$ MeV, the extra terms could modify the cross
section by about 30 percent or more.  For these two reactions we have
taken the full model of \cite{Doring:2005bx}.  We show now a band of values given by the uncertainties from the experimental helicity amplitudes of the $\gamma p\Delta^*$ transition \cite{Eidelman:2004wy} for the
cross section.  Furthermore, we show that both cross sections have been reduced
by about 30 percent with respect to those in \cite{Doring:2005bx} as a consequence of
the consideration of a more realistic $\Delta(1700)$ width, larger now and 
closer to the experimental 300 MeV, and in addition we have also taken a small
fraction of the $\rho N$ decay width in $d$-wave. These latter two cross
sections are also in qualitative agreement with preliminary results for the
cross sections as shown in \cite{nanovanstar}. 

   The  consideration  of the different reactions, with the cross
   sections spanning nearly two orders of magnitude, and the global qualitative
   agreement found for all the different reactions, gives support to the reaction
   mechanism suggested here in which a $\Delta^*(1700)$ resonance is excited which
   decays later on into $K \Sigma^*$ or $\eta \Delta$. The predictions of the
   cross sections are tied to the couplings of the $\Delta^*(1700)$ resonance  to
   the $K \Sigma^*$ or $\eta \Delta$ channels provided by the hypothesis that the
   $\Delta^*(1700)$ is a dynamically generated resonance. We showed in the 
   Introduction that the couplings to these channels were substantially
   different than those provided by a simple $SU(3)$ symmetry and there is hence
   substantial dynamical information from the underlying chiral dynamics and
   coupled-channels unitarity. In this respect the global analysis of these
   reactions offers support to the basic idea about the nature of the 
$\Delta^*(1700)$ as a dynamically generated resonance. No doubt a more detailed
theoretical analysis should consider extra terms, as done for instance in  
\cite{Doring:2005bx} and also extra mechanisms  beyond 200 or 300 MeV above the 
$\Delta^*(1700)$ region. However, the global qualitative agreement, considering
the large span of the different cross sections, and that a simple $SU(3)$
symmetrical consideration would produce cross sections about a factor 30
different, indicate that the agreement found here is not a trivial thing.  

At this point we would like to make some comments about other contributions to some of the reactions discussed and related ones that we mention below. First we mention that in the case of the $\gamma p\to K^+\pi^-\Sigma^+$ and $\gamma p\to K^+\pi^+\Sigma^-$ reactions there can be a contamination from the $K^+\Lambda(1405)$ production. This latter reaction was studied in \cite{Nacher:1998mi} and is presently investigated experimentally at Spring8/Osaka \cite{Ahn:2003mv}. We have recalculated the cross section for $\gamma p\to K^+\Lambda(1405)$ with the model of \cite{Nacher:1998mi} and find that the cross sections are of the same order of magnitude, or smaller in the case of the $\gamma p\to K^+\pi^+\Sigma^-$ reaction, than those shown in Fig. \ref{fig:photo}. Given the qualitative use made of the cross sections for these two reactions in the discussion above, and the large errors in the data, the conclusions drawn are not affected. Note also that for the $\gamma p\to K^0\pi^+\Lambda$ reaction the $\Lambda(1405)$ production is  not allowed and there the dominance of the $\Sigma^*(1385)$ production has stronger grounds. Note also that this latter reaction has a larger cross section than the other two from our $K\Sigma^*(1385)$ production mechanisms which should make potential extra contributions relatively smaller. 

Another possible mechanism can come from having the two mesons in a resonant state, the $K^*(892)$. Yet, the reactions $\gamma p(\pi p)\to \Delta^*(1700)\to K^*\Lambda$ are not allowed by isospin conservation, and precisely the reactions with $\Lambda$ in the final state are those with the largest cross sections. The sequence $\Delta^*(1700)\to K^*\Sigma$ is possible and this can be a source of background for the reactions with lower cross sections. Of course we could also have $\gamma p(\pi p)\to K^*\Lambda$ without passing through a $I=3/2$ channel and then this final state would also be allowed. However, note that the thresholds for the $K^*\Lambda$ and $K^*\Sigma$ production are $2007$ MeV and $2089$ MeV, respectively. Hence, for energies lower than these by 100 or 200 MeV this contribution should be highly suppressed. This is indeed an experimental fact as noted in \cite{curtis}. Yet, in some reactions the $K^*\Lambda$ contamination is more visible than in others, for instance in the $\pi^-p\to K^+\pi^-\Lambda$ reaction it is more apparent than for $\pi^-p\to K^0\pi^0\Lambda$ \cite{dahl}. This could explain why our model is short by about 30\% in the $\pi^-p\to K^+\pi^-\Lambda$ reaction cross section at 1930 MeV \cite{dahl} (not shown but also calculated) while it is good for the $\pi^-p\to K^0\pi^0\Lambda$ reaction. 

Similar comments can be made about the $\pi^-p\to K^+\pi^0\Sigma^-$ and $\pi^- p\to K^+\pi^-\Sigma^0$ reactions which in our model go via $K^+\Sigma^{*-}$ and, hence, can have $I=1/2,3/2$ while the $\pi^+p\to K^+\pi^+\Sigma^0$ and $\pi^+ p\to K^+\pi^0\Sigma^+$ reactions shown in Fig. \ref{fig:sigma} go through $K^+\Sigma^{*+}$ and, hence, have only $I=3/2$ and larger chances to couple to the \dell. The Clebsch-Gordan coefficients of $\pi^+ p\to\Delta^{*++}$ or $\pi^-p\to \Delta^{*0}$ also favor the $\pi^+p$ reactions and indeed we find cross sections for the $\pi^-p\to K^+\pi^0\Sigma^-$ and $\pi^-p\to K^+\pi^-\Sigma^0$ reactions substantially smaller  (about one order of magnitude). With such smaller cross sections it should be expected that background terms become more relevant and, hence, these reactions are not considered for our tests. 

Finally, let us mention the possible contribution in the entrance channels of other resonances, apart from the $\Delta^*(1700)$. The $N^*(1700)D_{13}$ can be a candidate which would possibly affect the $\pi^-p$ reactions but not the $\pi^+p$ reactions. In any case, the smaller width of the $N^*(1700)D_{13}$ (50-150 MeV) does not give much chance for contributions at the threshold of the reactions. Other possible $N^*$ or $\Delta^*$ resonances in the regions of the energies below 2000 MeV, considering their spin and parity, can be ruled out on the basis of the $s$-wave $K\Sigma^*$ dominance experimentally established in \cite{curtis,goussu}. The only possible exception is the one-star $\Delta(1940)D_{33}$ resonance for which no $K\Sigma^*$ or $\eta\Delta(1232)$ decay channels are reported in spite of being allowed by phase space and, hence, it is not considered here. 

The experimental $s$-wave $K\Sigma^*$ dominance at low energies served as support to our theory, but we should note that
at $\sqrt{s}=2020$ MeV in the $\pi^-p\to K^0\pi^0\Lambda$ reaction, the  $s$-wave $K\Sigma^*$ dominance from the $\sqrt{s}=1930$ data points \cite{curtis,goussu} does no longer hold; in fact, the production angular distribution shows a forward peak \cite{Thomas:1973uh} which is well described by $t$-channel $K^*$ exchange in the framework of the Stodolsky-Sakurai model \cite{Thomas:1973uh}. Similarly, the forward peak for the high energies is observed in the $\pi^+p\to K^+\pi^+\Lambda$ reaction \cite{Hanson:1972zz}. This does not rule out our production mechanisms since the Weinberg-Tomozawa term used to generate the \dell in the Bethe-Salpeter equation \cite{decu_ss} effectively accounts for a vector meson exchange in the $t$-channel (in the limit of small momentum transfer). Improvements could be done in the theory to account explicitly for the finite momentum transfer dependence, as done in \cite{Inoue:2001ip} but, since this only affects the larger energies, we do not consider it here.

\section{Conclusions}
We have looked at several pion-induced and photon-induced reactions at energies
above and close to the $\Delta^*(1700)$ with $K \pi \Lambda$, $K \pi \Sigma$ and 
$\eta \pi N$ in the final state.  We have made a theoretical model assuming that
a $\Delta^*(1700)$ is excited and then decays via the $K \Sigma^*(1385)$ or 
$\eta \Delta(1232)$ depending on the final state. We find that in spite of
exploiting the tail of the resonance, around half of the width or one width above
the nominal energy of the $\Delta^*(1700)$, the cross sections obtained are
sizable. The reason is the large couplings of the $\Delta^*(1700)$ to the 
$K \Sigma^*(1385)$ or $\eta \Delta(1232)$ which are provided by the theory in which
the $\Delta^*(1700)$ is a dynamically generated resonance. We showed that the
couplings squared to these channels were about 20-30 times bigger than estimated
by simple SU(3) symmetry arguments. We could also see that the presence of the 
 $ \Sigma^*(1385)$ was clear in the experimental data with 
 $K \pi \Lambda$, $K \pi \Sigma$  final states, with little room for background at low energies,
indicating a clear dominance of the   $ \pi \Lambda$, $ \pi \Sigma$ final states
in the $ \Sigma^*(1385)$ channel.  Despite the admitted room for improvements in
the theory, the qualitative global agreement of the different cross section 
with the data gives a strong support to the mechanisms proposed here and the
strong couplings of the $\Delta^*(1700)$ to the 
$K \Sigma^*(1385)$ or $\eta \Delta(1232)$ channels. The agreement found is more
significant when one realizes the large  difference in magnitude
of the different cross sections and the clear correlation of the theoretical
predictions with the data.

The results obtained are relevant because they rely upon the
$\Delta^*(1700)$ couplings to $K\Sigma^*$ and 
$\eta \Delta(1232)$ for which there is no
experimental information, but which are provided by the theory in which the
$\Delta^*(1700)$ is dynamically generated. 

The next question arises on what could be done in the future to make a more quantitative calculation. A number of factors is
needed to go forward in this direction: 
\begin{enumerate}
\item
The experimental total width of the $\Delta^*(1700)$ and the branching
ratio to $\pi N$ and $\rho N$ should be improved. The separation of the  
$\rho N$ channel in s- and d- waves needs also to be performed if accurate
predictions are to be done for 200-300 MeV above the $\Delta^*(1700)$
resonance region.
\item
Experiments at lower energies, closer to the $\Delta^*(1700)$ energy, 
for the photon- and pion-induced reactions discussed, would be most welcome.
There the $\Delta^*(1700)$ excitation mechanism would be more dominant 
and one would reduce uncertainties from other possible
background terms. 
\item                                           
Some improvements on the generation of the $\Delta^*(1700)$, including
extra channels to the $\pi\Delta$, $K\Sigma^*$, and $\eta\Delta$ used in
Refs. \cite{lutz,decu_ss}, like $\pi N$ in $d$-wave and $\rho N$ in $s
$- and $d$-waves, would be most welcome, thus helping fine tune the present
coupling of $\Delta^*(1700)$ to $K\Sigma^*$ and $\eta \Delta$ provided by 
\cite{decu_ss}.
\end{enumerate}
Awaiting progress in these directions, at the present time we could
claim, that within admitted theoretical and experimental uncertainties,
the present data for the large sample of pion and photon-induced reactions 
offer
support for the large coupling of the $\Delta^*(1700)$ resonance to $K
\Sigma^*$ and $\eta \Delta$ predicted by the chiral unitary approach for which
there was no previous experimental information.

\section*{Acknowledgments} This work is partly supported by DGICYT contract number
BFM2003-00856, and the E.U. EURIDICE network contract no. 
HPRN-CT-2002-00311. This
research is  part of the EU Integrated Infrastructure Initiative 
Hadron Physics Project
under  contract number RII3-CT-2004-506078.


\begin{thebibliography}{99}
\bibitem{dalitz}
R.H. Dalitz and S.F. Tuan, Ann. Phys. (N.Y.) 10, 307 (1960) 
\bibitem{Jennings:1986yg}
B.~K.~Jennings,
Phys.\ Lett.\ B {\bf 176}, 229 (1986).

\bibitem{Kaiser:1995cy}
N.~Kaiser, P.~B.~Siegel and W.~Weise,
Phys.\ Lett.\ B {\bf 362} (1995) 23

\bibitem{Kaiser:1996js}
N.~Kaiser, T.~Waas and W.~Weise,
Nucl.\ Phys.\ A {\bf 612} (1997) 297

\bibitem{angels}
E.~Oset and A.~Ramos,
Nucl. Phys. A {\bf 635} (1998) 99.

\bibitem{Nacher:1999vg}
J.~C.~Nacher, A.~Parreno, E.~Oset, A.~Ramos, A.~Hosaka and M.~Oka,
Nucl.\ Phys.\ A {\bf 678} (2000) 187

\bibitem{oller}
J.~A.~Oller and U.-G.~Mei{\ss}ner,
Phys.\ Lett.\ B {\bf 500} (2001) 263.

\bibitem{Inoue:2001ip}
T.~Inoue, E.~Oset and M.~J.~Vicente Vacas,
Phys.\ Rev.\ C {\bf 65} (2002) 035204

\bibitem{bennhold}
 E.~Oset, A.~Ramos and C.~Bennhold,
 Phys.\ Lett.\ B {\bf 527} (2002) 99
 [Erratum-ibid.\ B {\bf 530} (2002) 260]
 
\bibitem{Garcia-Recio:2002td}
  C.~Garcia-Recio, J.~Nieves, E.~Ruiz Arriola and M.~J.~Vicente Vacas,
  Phys.\ Rev.\ D {\bf 67} (2003) 076009
  
\bibitem{Jido:2003cb}
D.~Jido, J.~A.~Oller, E.~Oset, A.~Ramos and U.~G.~Meissner,
Nucl.\ Phys.\ A {\bf 725} (2003) 181

\bibitem{Garcia-Recio:2003ks}
C.~Garcia-Recio, M.~F.~M.~Lutz and J.~Nieves,
Phys.\ Lett.\ B {\bf 582} (2004) 49


\bibitem{Borasoy:2005ie}
  B.~Borasoy, R.~Nissler and W.~Weise,
  Eur.\ Phys.\ J.\ A {\bf 25}, 79 (2005)

\bibitem{Oller:2005ig}
  J.~A.~Oller, J.~Prades and M.~Verbeni,
  Phys.\ Rev.\ Lett.\  {\bf 95}, 172502 (2005)

\bibitem{Oller:2006jw}
  J.~A.~Oller,
  arXiv:hep-ph/0603134.


\bibitem{Prakhov:2004an}
  S.~Prakhov {\it et al.}  [Crystall Ball Collaboration],
  Phys.\ Rev.\ C {\bf 70}, 034605 (2004).
\bibitem{Eidelman:2004wy}
  S.~Eidelman {\it et al.}  [Particle Data Group],
  Phys.\ Lett.\ B {\bf 592}, 1 (2004).
\bibitem{Thomas:1973uh}
  D.~W.~Thomas, A.~Engler, H.~E.~Fisk and R.~W.~Kraemer,
  Nucl.\ Phys.\ B {\bf 56}, 15 (1973).




\bibitem{Magas:2005vu}
  V.~K.~Magas, E.~Oset and A.~Ramos,
  Phys.\ Rev.\ Lett.\  {\bf 95}, 052301 (2005)
  [arXiv:hep-ph/0503043].

\bibitem{lutz}
E.~E.~Kolomeitsev and M.~F.~M.~Lutz,
Phys.\ Lett.\ B {\bf 585} (2004) 243

\bibitem{decu_ss}
S.~Sarkar, E.~Oset and M.~J.~Vicente Vacas,
Nucl.\ Phys.\ A {\bf 750} (2005) 294

\bibitem{Sarkar:2005ap}
  S.~Sarkar, E.~Oset and M.~J.~Vicente Vacas,
  Phys.\ Rev.\ C {\bf 72}, 015206 (2005)
\bibitem{Roca:2006sz}
  L.~Roca, S.~Sarkar, V.~K.~Magas and E.~Oset,
  Phys.\ Rev.\ C {\bf 73}, 045208 (2006)



\bibitem{Doring:2006ub}
  M.~D\"oring, E.~Oset and S.~Sarkar,
  arXiv:nucl-th/0601027.


\bibitem{Doring:2005bx}
  M.~D\"oring, E.~Oset and D.~Strottman,
  Phys.\ Rev.\ C {\bf 73}, 045209 (2006), 
  arXiv:nucl-th/0510015.

\bibitem{nanovanstar}
M. Nanova at the ''International Workshop On The Physics Of Excited Baryons (NSTAR 05)'',
10-15 Oct 2005, Tallahassee, Florida
\bibitem{dahl}
O.~I.~Dahl, L.~M.~Hardy, R.~I.~Hess {\it et.al.}, Phys.\ Rev.\ {\bf 163}, 1337 (1967).
\bibitem{curtis}
L.~J.~Curtis, C.~T.~Coffin, D.~I.~Meyer, and K.~M.~Terwilliger, Phys.\ Rev.\ {\bf 132}, 1771 (1963).
\bibitem{Hanson:1972zz}
  P.~Hanson, G.~E.~Kalmus and J.~Louie,
  Phys.\ Rev.\ D {\bf 4}, 1296 (1971).

\bibitem{Grether:1973sz}
  D.~Grether, G.~Gidal and G.~Borreani,
  Phys.\ Rev.\ D {\bf 7}, 3200 (1973).


\bibitem{Hyodo:2003jw}
  T.~Hyodo, A.~Hosaka, E.~Oset, A.~Ramos and M.~J.~Vicente Vacas,
  Phys.\ Rev.\ C {\bf 68}, 065203 (2003)
\bibitem{Erbe:1970cq}
  R.~Erbe {\it et al.}  [Aachen-Berlin-Bonn-Hamburg-Heidelberg-Muenchen
                  Collaboration],
  Phys.\ Rev.\  {\bf 188}, 2060 (1969).
\bibitem{Erbe:2}
  R.~Erbe {\it et al.}  [Aachen-Berlin-Bonn-Hamburg-Heidelberg-Muenchen
                  Collaboration],
  Nuovo\ Cimento\ {\bf 49A}, 504 (1967).
\bibitem{cmc}
Cambridge Bubble Chamber Group, Phys.\ Rev.\ {\bf 156}, 1426 (1966)  


\bibitem{blatt}
J.M. Blatt and V.F. Weisskopff, {\it Theoretical Nuclear Physics} (Wiley, New York, 1952)

\bibitem{Manley:1984jz}
  D.~M.~Manley, R.~A.~Arndt, Y.~Goradia and V.~L.~Teplitz,
  Phys.\ Rev.\ D {\bf 30}, 904 (1984).

\bibitem{Manley:1992yb}
  D.~M.~Manley and E.~M.~Saleski,
  Phys.\ Rev.\ D {\bf 45}, 4002 (1992).

\bibitem{Butler:1992pn}
M.~N.~Butler, M.~J.~Savage and R.~P.~Springer,
Nucl.\ Phys.\ B {\bf 399}, 69 (1993)
\bibitem{Oset:2000eg}
E.~Oset and A.~Ramos,
Nucl.\ Phys.\ A {\bf 679}, 616 (2001)


\bibitem{Nacher:2000eq}
J.~C.~Nacher, E.~Oset, M.~J.~Vicente and L.~Roca,
Nucl.\ Phys.\ A {\bf 695}, 295 (2001)


\bibitem{Arndt:2003if}
R.~A.~Arndt, W.~J.~Briscoe, I.~I.~Strakovsky, R.~L.~Workman and M.~M.~Pavan,
Phys.\ Rev.\ C {\bf 69} (2004) 035213

\bibitem{Vrana:1999nt}
T.~P.~Vrana, S.~A.~Dytman and T.~S.~H.~Lee,
Phys.\ Rept.\  {\bf 328} (2000) 181

\bibitem{goussu}
O.~Goussu, M.~Sen\'e, B.~Ghidini {\it et.al.}, Nuovo Cimento {\bf 42A}, 606 (1966)
\bibitem{Nacher:1998mi} J.~C.~Nacher, E.~Oset, H.~Toki and A.~Ramos,
Phys.\ Lett.\ B {\bf 455} (1999) 55 
\bibitem{Ahn:2003mv} J.~K.~Ahn  [LEPS Collaboration],
Nucl.\ Phys.\ A {\bf 721} (2003) 715.
\end{thebibliography}
\end{document}